\documentclass[aps,prd,reprint,superscriptaddress,nofootinbib]{revtex4-1}
\pdfoutput=1
\usepackage{epsfig, subfigure}
\usepackage{setspace}
\usepackage{booktabs, tabularx} 
 
\usepackage{amsmath,bm,bbm}
\usepackage{color}
\usepackage{hyperref}
\usepackage{graphicx}
\usepackage{enumerate}
\hypersetup{
    pdfnewwindow=true,      
    colorlinks=true,       
    linkcolor=blue,          
    citecolor=blue,        
    filecolor=blue,      
    urlcolor=blue           
}

\def\lsim{\lesssim}  
\def\gsim{\gtrsim}

\newcommand{\beq}{\begin{equation}}  
\newcommand{\eeq}{\end{equation}}
\newcommand{\bea}{\begin{eqnarray}}  
\newcommand{\eea}{\end{eqnarray}}
\newcommand{\epm}{\ensuremath{e^{\pm}\;}}

\newcommand{\sigv}{\langle \sigma v\rangle}
\newcommand{\omh}{\Omega_{\rm CDM} h^{2}}

\newcommand{\ie}{{\it i.e.}}
\newcommand{\eg}{{\it e.g.}}
\newcommand{\etal}{{\it et al.}}

\begin{document}

\title{\mbox{\hspace{-0.5cm}Precise Relic WIMP Abundance and its Impact on Searches for Dark Matter Annihilation}}

\author{Gary Steigman}
\email{steigman.1@osu.edu}
\affiliation{\mbox{Center for Cosmology and AstroParticle Physics, Ohio State University, 191 W.~Woodruff Ave., Columbus, 43210 OH, USA}}
\affiliation{Dept. of Physics, Ohio State University, 191 W.~Woodruff Ave., Columbus,  43210 OH, USA}
\affiliation{Dept. of Astronomy, Ohio State University, 140 W.~18$^{th}$ Ave., Columbus,  43210 OH, USA}

\author{Basudeb Dasgupta}
 \email{dasgupta.10@osu.edu}
 \affiliation{\mbox{Center for Cosmology and AstroParticle Physics, Ohio State University, 191 W.~Woodruff Ave., Columbus, 43210 OH, USA}}

\author{John F.~Beacom}
\email{beacom.7@osu.edu}
\affiliation{\mbox{Center for Cosmology and AstroParticle Physics, Ohio State University, 191 W.~Woodruff Ave., Columbus, 43210 OH, USA}}
\affiliation{Dept. of Physics, Ohio State University, 191 W.~Woodruff Ave., Columbus,  43210 OH, USA}
\affiliation{Dept. of Astronomy, Ohio State University, 140 W.~18$^{th}$ Ave., Columbus,  43210 OH, USA}

\date{\today}

\begin{abstract}
If dark matter (DM) is a weakly interacting massive particle (WIMP) that is a thermal relic of the early Universe, then its total self-annihilation cross section is revealed by its present-day mass density. This result for a generic WIMP is usually stated as $\sigv\approx3\times10^{-26}\,{\rm cm^3s^{-1}}$, with unspecified uncertainty, and taken to be independent of WIMP mass. Recent searches for annihilation products of DM annihilation have just reached the sensitivity to exclude this canonical cross section for $100\,\%$ branching ratio to certain final states and small WIMP masses. The ultimate goal is to probe all kinematically allowed final states as a function of mass and, if all states are adequately excluded, set a lower limit to the WIMP mass. Probing the low-mass region is further motivated due to recent hints for a light WIMP in direct and indirect searches. We revisit the thermal relic abundance calculation for a generic WIMP and show that the required cross section can be calculated precisely. It varies significantly with mass at masses below $10\,{\rm GeV}$, reaching a maximum of $5.2\times10^{-26}\,{\rm cm^3s^{-1}}$ at $m\approx0.3\,{\rm GeV}$, and is $2.2\times10^{-26}\,{\rm cm^3s^{-1}}$ with feeble mass-dependence for masses above $10\,{\rm GeV}$. These results, which differ significantly from the canonical value and have not been taken into account in searches for annihilation products from generic WIMPs, have a noticeable impact on the interpretation of present limits from Fermi-LAT and WMAP+ACT.
\end{abstract}

\maketitle

\section{Introduction}                                                                 
\label{sec:introduction}
Cosmological measurements \cite{Komatsu:2010fb,Clowe:2006eq} have established that \mbox{$\sim80\,\%$} of the non-relativistic matter in the Universe is in the form of a non-luminous particle, dubbed ``Dark Matter'' (DM). Although there is no empirical evidence for a specific particle to be the DM, thermally populated weakly interacting massive particles~(WIMPs) are the best motivated candidates on theoretical grounds~\mbox{\cite{st85,Bertone:2004pz,Jungman:1995df}}. Particle physics theories addressing apparently unrelated issues, \eg, the hierarchy problem, often introduce new particles and a discrete symmetry that makes the least massive new particle stable. This provides a DM candidate. In such theories, the observed cosmological abundance of DM can be explained by the chemical ``freeze out'' of a thermal relic~\cite{zeld65, Chiu:1966kg, Lee:1977ua,Hut:1977zn,Wolfram:1978gp,Steigman:1979kw,Bernstein:1985th,Scherrer:1985zt,Srednicki:1988ce,Gondolo:1990dk}. See also reviews~in~\cite{Bertone:2004pz, Jungman:1995df, Steigman:1979kw, Kolb:1990vq}.

The observationally inferred relic abundance of DM is a valuable empirical clue to the particle nature of the WIMP. The interactions that determine the relic abundance of DM in the Universe also lead to annihilation of DM pairs to other particles in the present epoch. The aim of indirect detection experiments is to observe a flux of the annihilation products created in astrophysical environments where DM annihilation may be occurring at an appreciable rate. Similarly, collider experiments are attempting to produce WIMPs with these cross sections. Unambiguous detection of DM annihilation has not been achieved yet, but several experiments are now probing annihilation cross sections that are expected of a thermal WIMP. The relationship of the annihilation cross section to the cross section for scattering on other particles is well-defined but model dependent. Direct detection experiments search for nuclear recoils resulting from the scattering of local DM (from the DM population of the Milky Way halo). 

Our aim in this work is a contemporary reappraisal of the thermal WIMP relic abundance and its relationship to constraints from indirect detection experiments. This is motivated and timely for a number of reasons, as we discuss below.

First, although the calculation relating the thermal relic abundance and the annihilation cross section has been done several times in the literature~\cite{zeld65, Chiu:1966kg, Lee:1977ua,Hut:1977zn,Wolfram:1978gp,Steigman:1979kw,Bernstein:1985th,Scherrer:1985zt,Srednicki:1988ce,Gondolo:1990dk}, we identify several simple improvements to the analytical approach, that can be made quite easily. Furthermore, the original relic density calculations were performed in an era when the target DM relic density and the evolution with temperature of the early Universe radiation density were rather uncertain. As a result, the required annihilation cross section could not be predicted with much precision. The situation has changed dramatically since then. The cosmological and particle physics inputs are now determined much more accurately so that the required annihilation cross section can be predicted more precisely. In particular, numerical routines can perform the required calculations to high precision, \eg, {\tt DarkSUSY}~\cite{Gondolo:2004sc} for supersymmetric models, and {\tt micrOMEGAs}~\cite{Belanger:2006is} for a wider variety of models. However, considering that nowadays there is an increased interest in generic dark matter candidates, we believe that a precise analytical calculation, revealing how the result depends on the WIMP mass and the present mass density, is of value. Our work, with updated inputs, improvements in the analytic calculation, comparison between the analytic and numerical calculations, including a careful discussion of the errors resulting from the analytic approximations, and extension to lower masses, will be useful for testing generic WIMP models.

Second, experiments are now probing annihilation cross sections close to the ``canonical" value of $\sigv = 3\times 10^{-26}\,{\rm cm^{3}s^{-1}}$. Studies, \eg, from Fermi-LAT gamma-ray data from nearby dwarf galaxies and diffuse emission in our galaxy, have in fact constrained cross sections to be lower than this canonical thermal cross section, and disfavor WIMPs for a range of DM masses for annihilations to $b\bar{b}$ and $\tau\bar{\tau}$~\cite{Abdo:2010dk, Abazajian:2010sq, GeringerSameth:2011iw, collaboration:2011wa}. Studies of galaxy clusters are also probing this range of cross sections, albeit with greater uncertainties associated with modeling the cluster DM halo~\cite{Ackermann:2010rg, Ando:2012vu, Han:2012au}. We find, both numerically and analytically, that for small masses, $\sigv$ is smaller than the canonical value assumed in previous studies, weakening the claimed mass limits by up to a factor of two. Future experiments will continue to probe this interesting range of cross sections more aggressively and for other channels, making this present study timely.

Third, a variety of recent experiments have made WIMP masses $\lsim10\,$GeV very interesting~\cite{Hooper:2012ft}. Hints from direct detection experiments, \eg, DAMA/LIBRA, CoGeNT, CRESST-II etc.~\cite{Bernabei:2010mq, Aalseth:2011wp,Aalseth:2010vx,Angloher:2011uu}, relate to the annihilation cross section indirectly, in a model dependent fashion, while the astrophysical observations directly probe the annihilation cross section~\cite{Finkbeiner:2004us,Hooper:2007kb,Hooper:2010mq,Hooper:2010im,Linden:2011au,Hooper:2011ti}. Particular attention should be focused on the value of the DM annihilation cross section for this range of low WIMP masses. For this regime of small WIMP masses, which has traditionally neither been favored nor investigated in the previous literature, we find a factor of $\gsim 2$ increase in the value of $\sigv$ required to achieve the target WIMP relic density. As a result, the present day annihilation fluxes are increased by the same factor for these low masses.

Exploring this low mass regime is also timely because low mass thermal relics are in tension with cosmological constraints on reionization and recombination from WMAP+ACT CMB observations~\mbox{\cite{Cirelli:2009bb,Hutsi:2011vx,Galli:2011rz}}. It should be noted that these constraints require significant modeling and may be evaded, \eg, by annihilation predominantly to neutrinos, making the limits significantly weaker~\mbox{\cite{Beacom:2006tt, Yuksel:2007ac, Yuan:2010gn, Abbasi:2011eq}}. However, if the CMB constraints do apply, our results complement and strengthen them.

Motivated by the above considerations, in this paper we revisit the relic abundance calculation for the simplest WIMP model, adopting the standard model (SM) particle spectrum to which we add one additional Majorana fermion, the DM candidate, that self-annihilates via s-wave scattering.  The relic abundance calculation is reviewed with particular attention paid to the inputs and assumptions and we investigate their impact of their uncertainties on the final result. We show that since most inputs are measured to much better than $10\,\%$ precision, the required annihilation cross section can be predicted with $\lsim {\rm few}\,\%$ uncertainty. We focus on the lower WIMP masses and find that for $m\lsim 10\,$GeV, the required cross section increases with decreasing mass, rising to as much as $\sigv \sim 5.2\times10^{-26}\,{\rm cm^3s^{-1}}$. In contrast, for larger masses, $\gsim 15\,{\rm GeV}$, we find that over some four orders of magnitude in mass, the required cross section is roughly constant (increasing logarithmically) to within $\lsim 7\,\%$, at a value of \mbox{$\sigv = 2.2\times10^{-26}\,{\rm cm^3s^{-1}}$}, \ie, a value which is $\sim 40\,\%$ smaller than the canonical value, $\sigv = 3\times10^{-26}\,{\rm cm^3s^{-1}}$, quoted extensively in the literature. These differences, which have never been taken into account for generic WIMPs, now impact the Fermi-LAT limits to the WIMP annihilation cross section and the lower bound to the WIMP mass derived from them~\cite{GeringerSameth:2011iw, collaboration:2011wa}, and they strengthen the cosmological WMAP/ACT constraints~\mbox{\cite{Cirelli:2009bb,Hutsi:2011vx,Galli:2011rz}}.

The outline of this paper is as follows. An improved relic abundance calculation is presented in \S\,\ref{sec:analytical} where we derive an approximate analytical expression relating the relic abundance $\omh$ to the WIMP annihilation cross section $\sigv$ as a function of the WIMP mass, identifying and including terms which might affect the result at the $\sim 1\,\%$ level.  In \S\,\ref{sec:numerical}, we verify the approximate analytic results against those from a direct numerical integration of the evolution equation, confirming that the analytical results are accurate to $\sim3\,\%$ or better. In \S\,\ref{sec:limits} we compare the total annihilation cross section required to produce the cold dark matter abundance inferred from the WMAP-7 observations~\cite{Komatsu:2010fb}, as a function of the WIMP mass, to the limits on the partial cross sections for annihilation into particular channels derived from the Fermi-LAT and WMAP+ACT data and, we show the impact of our new results on the derived mass limits. In \S\,\ref{sec:summary}, we summarize our conclusions and provide an outlook to future developments.

\section{Evolution Of Thermal Relics In The Early Universe}
\subsection{Review of the Framework}
\label{sec:framework}

We consider a stable WIMP $\chi$ of mass $m$, produced thermally during the early evolution of the Universe, and follow its evolution as the Universe expands and cools. For concreteness, we take $\chi$ to be a spin 1/2 Majorana fermion, so that $\chi$ is its own antiparticle and has $g_{\chi}=2$ degrees of freedom.  We assume these particles are sufficiently coupled to the photons and the other particles present in the early Universe so that they are produced by the relativistic plasma and establish a common temperature $T (\equiv T_{\gamma})$ with it. Their evolution, determined by the competition between production and annihilation, is described by\,\cite{zeld65},
\beq
{dn \over dt} + 3Hn=\frac{d(na^3)}{a^3 dt}=\sigv\left(n_{eq}^2-n^2\right)\,,
\label{eq:evol1}
\eeq
where $n$ is the number density of $\chi$'s, $a$ is the cosmological scale factor, the Hubble parameter $H=a^{-1}da/dt$ provides a measure of the universal expansion rate, and $\sigv$ is the thermally averaged annihilation rate factor (``cross section''). For the most part we use natural units with $\hbar\equiv c\equiv k \equiv 1$.  When $\chi$ is extremely relativistic~\mbox{($T \gg m$)}, the equilibrium density $n_{eq} = 3\zeta(3)g_{\chi}T^3/(4\pi^2)$, where $\zeta(3)\approx1.202$.  In contrast, when $\chi$ is non-relativistic~\mbox{($T \lsim m$)}, its equilibrium abundance is \mbox{$n_{eq}=g_{\chi} \left(mT/(2\pi)\right)^{3/2}\exp(-m/T)$}.  If $\chi$ could be maintained in equilibrium, $n = n_{eq}$ and its abundance would decrease exponentially.  However, when the $\chi$ abundance becomes very small, equilibrium can no longer be maintained (the $\chi$'s are so rare they can't find each other to annihilate) and their abundance freezes out.  This process is described next.

\begin{figure}[!t]
\includegraphics[width=\columnwidth]{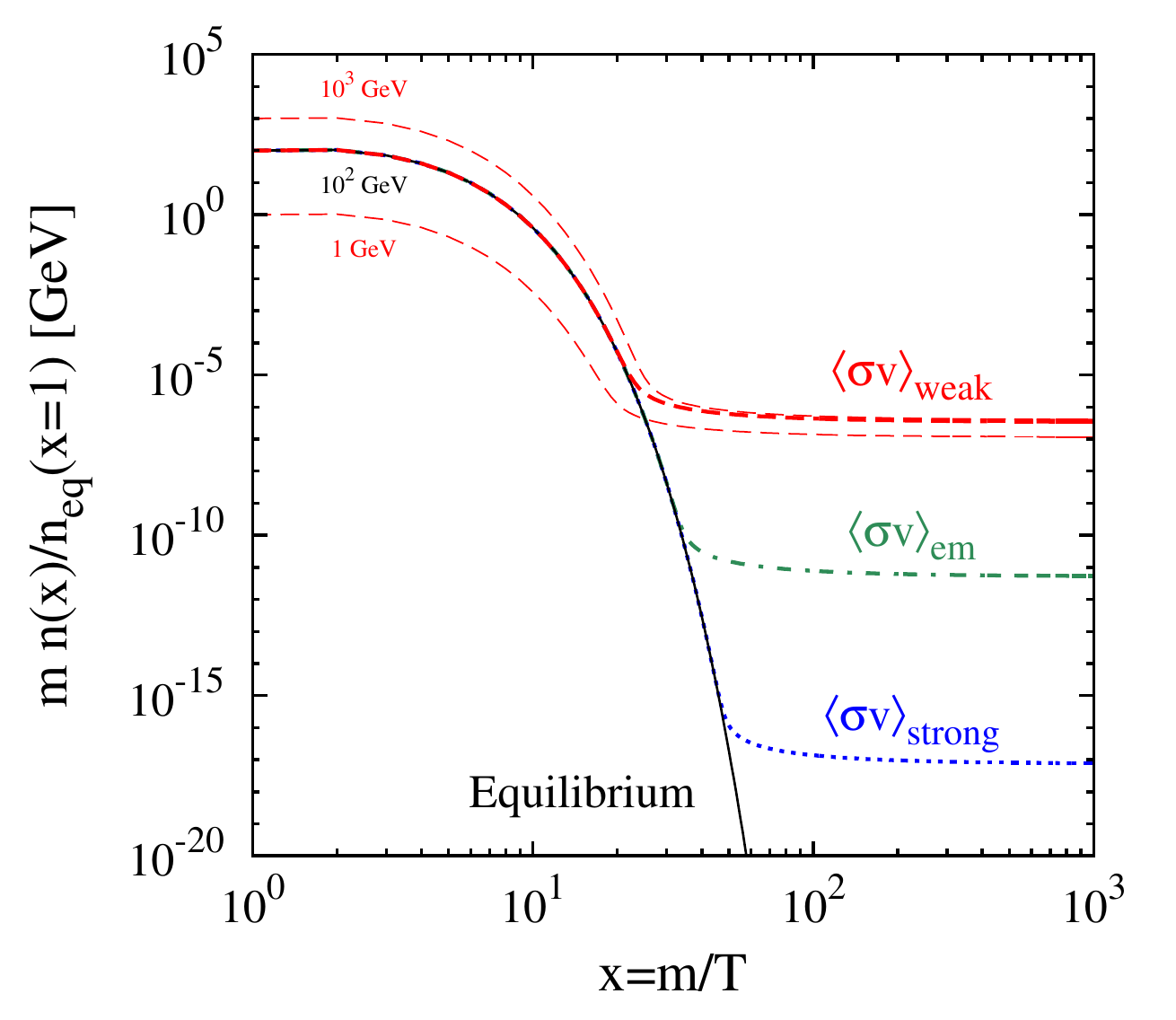}
\caption{Evolution of the cosmological WIMP abundance as a function of $x = m/T$. Note that the y-axis spans 25 orders of magnitude. The thick curves show the WIMP mass density, normalized to the initial equilibrium number density, for different choices of annihilation cross section $\sigv$ and mass $m$. Results for $m=100\,{\rm GeV}$, are shown for weak interactions, $\sigv=2\times10^{-26}\,{\rm cm^3s^{-1}}$, (dashed red), electromagnetic interactions, $\sigv=2\times10^{-21}\,{\rm cm^3s^{-1}}$ (dot-dashed green), and strong interactions, $\sigv=2\times10^{-15}\,{\rm cm^3s^{-1}}$ (dotted blue). For the weak cross section the thin dashed curves show the WIMP mass dependence for $m=10^3\,$GeV (upper dashed curve) and $m=1\,$GeV (lower dashed curve). The solid black curve shows the evolution of the equilibrium abundance for $m=100\,$GeV. This figure is an updated version of the figure which first appeared in Steigman (1979)~\cite{Steigman:1979kw}.}
\label{fig:evol}
\end{figure}

We begin by referring to Fig.\,\ref{fig:evol}, where the evolution of the mass density of WIMPs of mass $m$, normalized to the initial equilibrium WIMP number density, is shown as a function of $x=m/T$, which is a proxy for ``time'', for different values of $\sigv$. With this definition, the final asymptotic value is proportional to the relic abundance, as will be seen later.  Later in this section it is explained how this evolution is calculated, but first we call attention to some important features.  During the early evolution when the WIMP is relativistic ($T\gsim m$), the production and annihilation rates far exceed the expansion rate and $n = n_{eq}$ is a very accurate, approximate solution to Eq.\,(\ref{eq:evol1}).  It can be seen in Fig.\,\ref{fig:evol} that, even for $T \lesssim m$, the actual WIMP number density closely tracks the equilibrium number density (solid black curve).  As the Universe expands and cools and $T$ drops further below $m$, WIMP production is exponentially suppressed, as is apparent from the rapid drop in $n_{eq}$. Annihilations continue to take place at a lowered rate because of the exponentially falling production rate.  At this point, equilibrium can no longer be maintained and, $n$ deviates from (exceeds) $n_{eq}$. However, even for $T \lsim m$, the annihilation rate is still very fast compared to the expansion rate and $n$ continues to decrease, but more slowly than $n_{eq}$.  For some value of $T \ll m$, WIMPs become so rare that residual annihilations also cease and their number in a comoving volume stops evolving (they ``freeze out"), leaving behind a thermal relic.

It is well known that weak-scale cross sections naturally reproduce the correct relic abundance in the Universe, whereas other stronger (or weaker) interactions do not. This is a major motivation for WIMP dark matter. Note that while for ``high" masses (\mbox{$m \gsim 10$~GeV}) the relic abundance is insensitive to $m$, for lower masses the relic abundance depends sensitively on mass, increasing (for the same value of $\sigv$) by a factor of two.

There are two clearly separated regimes in this evolution -- ``early" and ``late". The evolution equation (Eq.\,(\ref{eq:evol1})) can be solved analytically by different approximations in these two regimes.  During the early evolution, when the actual abundance tracks the equilibrium abundance very closely ($n \approx n_{eq})$, the rate of departure from equilibrium, $d(n - n_{eq})/dt$, is much smaller than the rate of change of $dn_{eq}/dt$.  In the late phase, where $n\gg n_{eq}$, the equilibrium density $n_{eq}$ may be ignored compared to $n$ and Eq.\,(\ref{eq:evol1}) may be integrated directly. This strategy allows the evolution to be solved analytically in each of the two regimes and then joined at an intermediate matching point which we call $x_{*}$.  Because the deviation from equilibrium, $(n - n_{eq})$, is growing exponentially for $x \approx x_{*}$, the value of $x_{*}$ is relatively insensitive (logarithmically sensitive) to the choice of $(n - n_{eq})_{*}$.

Since the dynamics leading to freeze out occurs during the early, radiation dominated ($\rho = \rho_{\rm R}$) evolution of the Universe, it is useful to recast physical quantities in terms of the cosmic background radiation photons. The total radiation density may be written in terms of the photon energy density ($\rho_{\gamma}$) as $\rho = (g_{\rho}/g_{\gamma})\rho_{\gamma}$ where, $g_\rho$ counts the relativistic ($m < T$) degrees of freedom contributing to the energy density,
\beq
g_{\rho} \equiv \sum_{\rm B}~g_{\rm B}\bigg({T_{\rm B} \over T_{\gamma}}\bigg)^{4} + {7 \over 8}\sum_{\rm F}~g_{\rm F}\bigg({T_{\rm F} \over T_{\gamma}}\bigg)^{4}\,.
\label{eq:gdef}
\eeq
In Eq.\,(\ref{eq:gdef}), ${\rm B} \equiv$~Bosons and ${\rm F} \equiv$~Fermions.  For those particles in thermodynamic equilibrium with the photons, $T_{\rm B,F} = T_{\gamma}$ (in the following we drop the subscript $\gamma$ and write $T \equiv T_{\gamma}$).  If relativistic particles are present that have decoupled from the photons, it is necessary to distinguish between two kinds of $g$: $g_{\rho}$ in Eq.\,(\ref{eq:gdef}) is associated with the total energy density, whereas $g_{s}$ is associated with the total entropy density,
\beq
g_{s} \equiv \sum_{\rm B}~g_{\rm B}\bigg({T_{\rm B} \over T_{\gamma}}\bigg)^{3} + {7 \over 8}\sum_{\rm F}~g_{\rm F}\bigg({T_{\rm B} \over T_{\gamma}}\bigg)^{3}.
\label{eq:gsdef}
\eeq 
Note that $g_{\rho}$ and $g_{s}$ differ only when there are relativistic particles present that are not in equilibrium with the photons, \ie, when $T_{\rm B,F} \neq T_{\gamma}$. For the SM particle content this only occurs for $T \lsim m_{e}\,$, when the \epm pairs annihilate, heating the photons relative to the neutrinos ($T_{\gamma} > T_{\nu}$), after the neutrinos have decoupled ($T_{\nu,dec} \sim 2-3$~MeV). 

In the absence of phase transitions, throughout the evolution of the Universe the entropy in a comoving volume, $S \equiv sa^3 = (2\pi^2/45)g_{s}T^{3}a^3$, is conserved. As a result, in the evolution equation, Eq.\,(\ref{eq:evol1}), $n$ may be replaced with $Y \equiv n/s$ and $n_{eq}$ with $Y_{eq} = n_{eq}/s$ where, 
\beq
Y_{eq} = {n_{eq} \over s}={45\over 2\pi^4}\bigg({\pi\over 8}\bigg)^{1/2}{g_\chi \over g_s}\,x^{3/2}{\rm exp}(-x).
\eeq
Entropy conservation also enables us to relate changes in the scale factor and the temperature\footnote{As an aside, we note that the total entropy in a comoving volume is $S = 1.80g_{s}N_{\gamma}$, where $N_{\gamma}$ is the number of photons in the comoving volume.  As the Universe expands and cools, $dS = 0$, so that the ``photon evolution equation" is $d(g_{s}N_{\gamma})/dt= 0$,  and $g_{s}(T)N_{\gamma}(T) =$~constant.  This reflects the fact that as the temperature drops below the masses of the SM particles in thermal equilibrium with the photons, they annihilate and/or decay, ``heating" the photons (\ie, creating more photons in the comoving volume). The temperature is always a monotonically decreasing function of time or scale factor, but $T$ decreases more slowly than $1/a$.  This result is not unique to photons; it applies to all extremely relativistic particles in thermal equilibrium with the photons.}.  During radiation dominated epochs the expansion rate of the Universe ($H$) is related to the total energy density by $H\equiv (1/a)da/dt = \sqrt{8\pi G\rho/3}$, where $\rho = (\pi^2/30)g_{\rho}T^{4}$. Following the evolution using $x\equiv m/T$ instead of $t$,
\beq
{dY \over dx} = {s\sigv \over Hx}\bigg[1 + {1 \over 3}{d({\rm ln}g_{s}) \over \,d({\rm ln}T)}\bigg](Y_{eq}^{2} - Y^{2}).
\label{eq:workingeq}
\eeq
In Eq.\,(\ref{eq:workingeq}) the term in square brackets accounts for the variation of $g_{s}$ with $T$ (as in~\cite{Srednicki:1988ce}), an effect which is almost always neglected.  Equation\,(\ref{eq:workingeq}) is our starting point for both analytical and numerical investigations. 

\begin{figure}[!t]
\includegraphics[width=1.0\columnwidth]{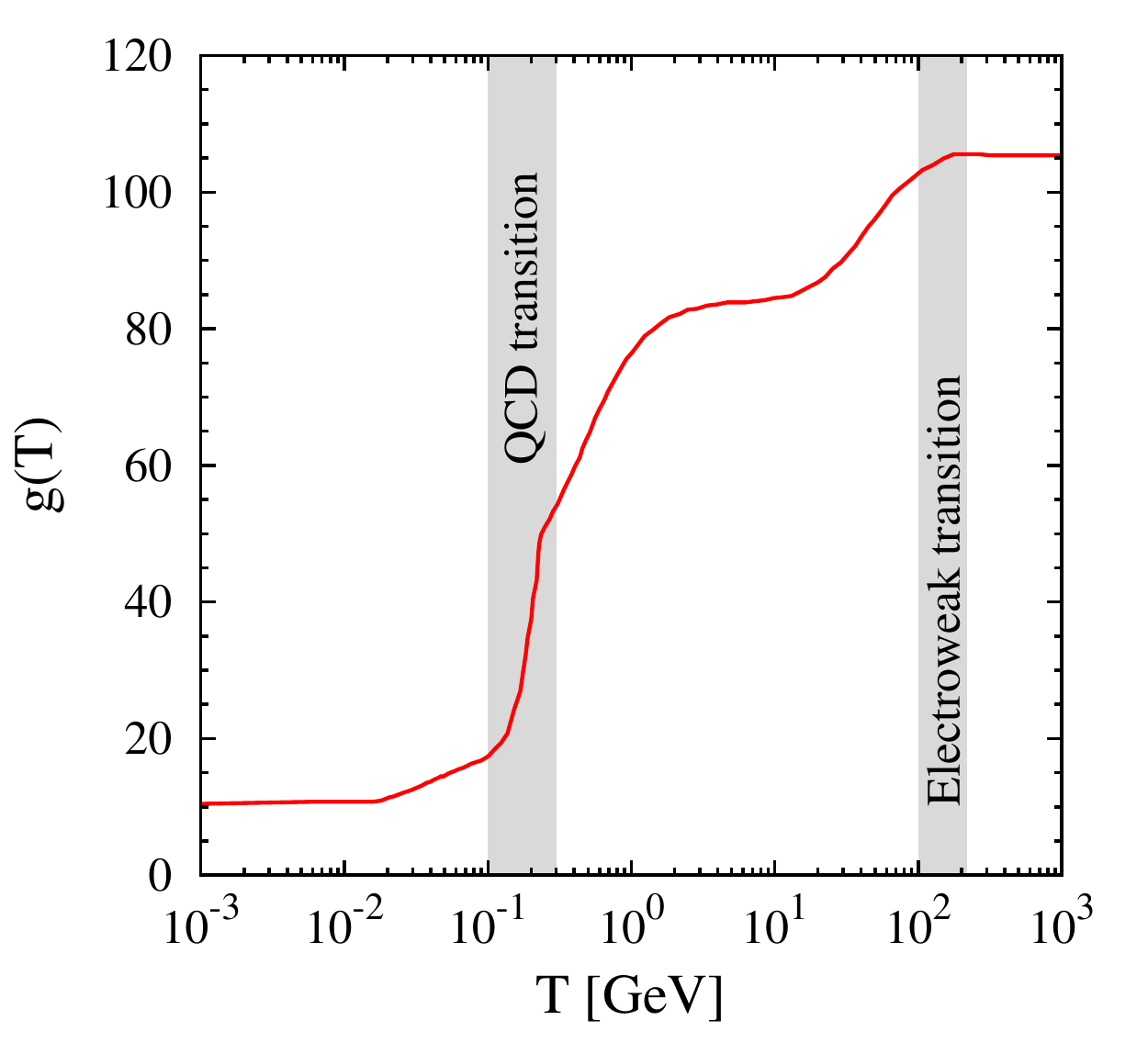}
\caption{The effective number of interacting (thermally coupled), relativistic degrees of freedom, $g$, as a function of the temperature for $1\,{\rm MeV} \leq T \leq 1\,{\rm TeV}$ (adapted from Laine and Schroeder~\cite{Laine:2006cp}).}
\label{fig:gls}
\end{figure}

Equation\,(\ref{eq:workingeq}) makes it clear that the only source of uncertainty and model dependence in this calculation is from $g(T)$, which enters directly into $s \propto g_{s}T^{3}$ and~\mbox{$H \propto g_{\rho}^{1/2}T^{2}$}.  For our calculations  we use $g=g_{\rho}=g_{s}$ because for the range of WIMP masses we consider, $10\,{\rm MeV}\lsim m\lsim 10\,{\rm TeV}$, and the particle content of the SM, $T_{\rm B} = T_{\rm F} = T$, so there is no distinction between $g_{\rho}$ and $g_{s}$. We adopt $g(T)$ from the calculations of Laine and Schroeder~\cite{Laine:2006cp}; $g$ as a function of $T$ is shown in Fig.\,\ref{fig:gls} for temperatures in the range $1~{\rm MeV} \leq T \leq 1~{\rm TeV}$.  Over this range of six orders of magnitude in temperature, $g$ changes only by a factor of $\sim10$.  The relatively rapid rise in $g$ versus $T$ for the temperature interval $0.1\,{\rm GeV} \lsim T \lsim 1\,{\rm GeV}$ reflects the quark-hadron transition, which is a crossover transition and not a phase transition~\mbox{\cite{Brown:1988qe, Aoki:2006we, Gupta:2011wh}}. These results are expected to be accurate to within a few\,$\,\%$ everywhere, except in the region of the quark-hadron transition and electroweak transition (ignored here), where the errors are expected to be~$\lsim10\,\%$~\cite{privateLS}.

The strategy is to assume that the WIMP begins in equilibrium for $x\gsim1$ and to solve Eq.\,(\ref{eq:workingeq}) for $x \rightarrow \infty$ ($x \gg 1$), to find $Y$ at present ($t = t_{0},\ T = T_{0}$). The present relic abundance, $\rho_{\chi}$, may be written in terms of the density parameter $\Omega$ and the critical mass density~$\rho_{crit}$,
\beq
\Omega \equiv \rho_{\chi}/\rho_{crit},
\eeq
where
\beq
\rho_{\chi}=m\,s_0Y_0,
\label{eq:rhochidef}
\eeq
and $8\pi G\rho_{crit}=3H_0^2$. The subscript $0$ denotes quantities evaluated at the present time, when $T_0=2.725\pm0.001\,$K~\cite{pdg}. For a given WIMP mass, this allows us to find the value of $\sigv$ required in order to match, \eg, the \mbox{WMAP-7} inferred result, $\Omega_{\rm CDM}h^2=0.1120\pm0.0056$~\cite{Komatsu:2010fb}.

\subsection{Improved Analytical Treatment}
\label{sec:analytical}

We first solve Eq.\,(\ref{eq:workingeq}) using an analytical approximation (see, \eg, \cite{zeld65}). Although our calculation mirrors those in the previous literature, we improve upon earlier results by carefully including the effect of the changing value of $g(T)$ during the evolution, as well as by including some terms ignored in previous studies.

\subsubsection{Early Evolution ($n \approx n_{eq}$)}
\label{sec:approx1}

In following the early evolution where $Y \approx Y_{eq}$, it is useful to write \mbox{$Y \equiv (1 + \Delta)Y_{eq}$} and to follow the evolution of $\Delta$ instead of $Y$.  The evolution equation for $\Delta$ takes the form
\begin{multline}
{d({\rm ln}(1+\Delta)) \over d({\rm ln}x)} + {d({\rm ln}Y_{eq}) \over d({\rm ln}x)} = \\ 
-{\Gamma_{eq}\over H}\bigg[1 + {1 \over 3}{d({\rm ln}\,g) \over d({\rm ln}\,T)}\bigg]{\Delta(2+\Delta) \over (1+\Delta)},
\label{eq:logevolution}
\end{multline}
where $\Gamma_{eq} \equiv n_{eq}\sigv=Y_{eq}s\sigv$ and 
\beq
\Gamma_{eq}/H = 8.00\times 10^{34}m\sigv x^{1/2}e^{-x}g^{-1/2}\,,
\label{eq:GbyHdef}
\eeq
where $m$ is in GeV, and $\sigv$ is in ${\rm cm^3s^{-1}}$. We use these units throughout this section, and wherever it is unstated, this should be assumed. Now, since \mbox{$Y_{eq} \propto x^{3/2}e^{-x}/g$},
\beq
{d({\rm ln}Y_{eq}) \over d({\rm ln}\,x)} = -\bigg[x-3/2 + {d({\rm ln}\,g) \over d({\rm ln}\,x)}\bigg].
\eeq
This allows us to rearrange Eq.\,(\ref{eq:logevolution}) as 
\begin{multline}
{\Delta(2+\Delta) \over (1+\Delta)} = {{x-3/2 - {\dfrac{d({\rm ln}\,g)}{d({\rm ln}T)}} - {\dfrac{d({\rm ln}(1+\Delta))}{d({\rm ln}\,x)}}}\over{\dfrac{\Gamma_{eq}}{H}\bigg[1 + \dfrac{1}{3}\dfrac{d({\rm ln}\,g)}{d({\rm ln}\,T)}\bigg]}}\,. 
\label{eq:deltagen}
\end{multline}
Note that although the logarithmic derivative of $g$ with respect to $T$ in the denominator on the right hand side has been noted before~\cite{Srednicki:1988ce}, the third term in the numerator, involving the same derivative, has not been considered in previous treatments. If freeze out occurs in a temperature regime where $g$ is changing, both of these terms are equally important.

If the WIMP is close to equilibrium, \ie, $\Delta,\,d\Delta/dx\ll1$,  the fourth term in the numerator of Eq.\,(\ref{eq:deltagen}) can be ignored\footnote{Since $\Delta$ is increasing exponentially, this neglect becomes a poor approximation when $\Delta \gsim {\cal O}(1)$.}. If, further, the terms involving the logarithmic derivative of $g$ with $T$ are ignored and Eq.\,(\ref{eq:GbyHdef}) is used,
\beq
{\Delta(2+\Delta) \over (1+\Delta)} \approx {1.25\times 10^{-35}g^{1/2} \over \sigv m}\bigg({(x-3/2)e^{x} \over x^{1/2}}\bigg)\,.
\label{eq:delta}
\eeq
Comparison with the results from the numerical integration of the evolution equation confirms that the neglect of the logarithmic derivative of $g$ introduces an error which is $< 1\,\%$, except when the approach to freeze out occurs close to the quark hadron transition. As in almost all previous analytic analyses, it can be assumed that $(x - 3/2)/x^{1/2} \approx x^{1/2}$, introducing a very small error of order $\sim 0.1 - 1\,\%$. 

\begin{figure}[!t]
\includegraphics[width=1.0\columnwidth]{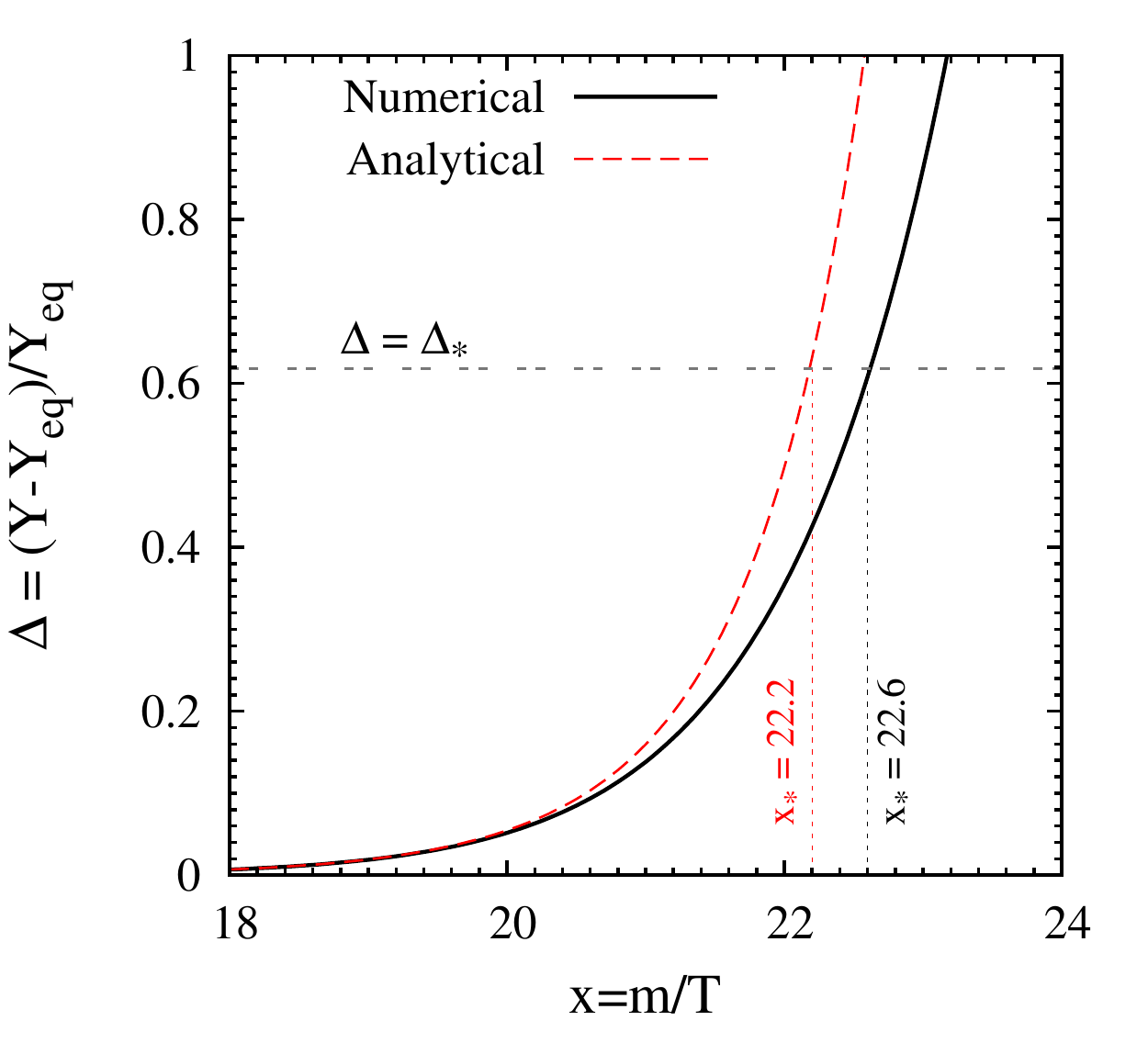}
\caption{Evolution of the departure of the WIMP abundance from the equilibrium abundance, $\Delta$, for $x$ close to $x_*$. The departure from the equilibrium value is shown as a function of $x$, calculated numerically (solid black), and analytically (dashed red) using Eq.\,(\ref{eq:delta}), for an illustrative case with $m=100\,$GeV and $\sigv=2.2\times10^{-26}\,{\rm cm^3s^{-1}}$. The analytical approximation ignores $d\Delta/dx$ (see Eq.\,(\ref{eq:deltagen})), leading to an underestimate of $x_*$ by $\sim 2\,\%$. See the text for details.}
\label{fig:approx}
\end{figure}

The departure from equilibrium, $\Delta$, is shown as a function of $x$ in Fig.\,\ref{fig:approx} for an illustrative case with $m=100\,$GeV and $\sigv=2.2\times10^{-26}\,{\rm cm^3s^{-1}}$. The numerically calculated value (solid black curve) is lower than the analytical prediction using Eq.\,(\ref{eq:delta}) (dashed red curve).  This is because the analytical approximation ignores $d\Delta/dx$ in Eq.\,(\ref{eq:deltagen}), which is not completely negligible. As a result the analytical prediction for $\Delta$ (dashed red curve in Fig.\,\ref{fig:approx}) overshoots the true value (solid black curve in Fig.\,\ref{fig:approx}), leading to an underestimate of $x_*$ by about $2\,\%$.

$\Delta$ is initially very small, but, as may be seen from Fig.\,\ref{fig:approx}, as $x$ increases, $\Delta$ increases exponentially, eventually approaching ${\cal O}(1)$. Beyond this point the approximations ($\Delta,d\Delta/dx\ll1$) leading to Eq.\,(\ref{eq:delta}) break down. Therefore, the above analysis can only be valid for $x\lsim x_{*}$, where $\Delta \lsim {\cal O}(1)$.  We define $x_{*}$ by setting the left hand side of Eq.\,(\ref{eq:delta}) to 1 when $x = x_{*}$,
\beq
{\Delta(x_*)\left(2+\Delta(x_*)\right) \over \left(1+\Delta(x_*)\right)}= 1\,,
\eeq
resulting in $\Delta_*\equiv\Delta(x_*)=(\sqrt{5}-1)/2\approx0.618$. Our results for $T_{*}$ ($x_{*}$) and for $\sigv$ depend logarithmically on this choice of $\Delta_{*}$. Some alternative choices are $\Delta_{*} = 1/2$ or $\Delta_{*} = \sqrt{2} - 1$.  We explicitly verified that these alternate choices would change our result for \mbox{$\sigv$~by~$\sim\pm 0.1\%\,$}. 

The solution for $x_{*}$ from Eq.\,(\ref{eq:delta}) for $\Delta_{*} = 0.618$ is
\begin{multline}
x_{*} + {\rm ln}(x_{*} - 1.5) - 0.5\,{\rm ln}\,x_{*} = \\ 
20.5 + {\rm ln}(10^{26}\sigv) + {\rm ln}\,m - 0.5\,{\rm ln}\,g_{*}.
\label{eq:x*vsm}
\end{multline}
This equation is solved iteratively for $x_{*}$ as a function of the WIMP mass $m$ (in GeV), $\sigv$, and $g_{*}$. If $T_{*} = m/x_{*}$ is close to the region where $d({\rm ln}g)/d({\rm lnT})\sim1$, \eg, close to the temperature of the quark hadron transition, Eq.\,(\ref{eq:deltagen}) can be solved iteratively (with $d\Delta/dx=0$), for a more accurate result.  

\begin{figure}[!t]
\includegraphics[width=0.92\columnwidth]{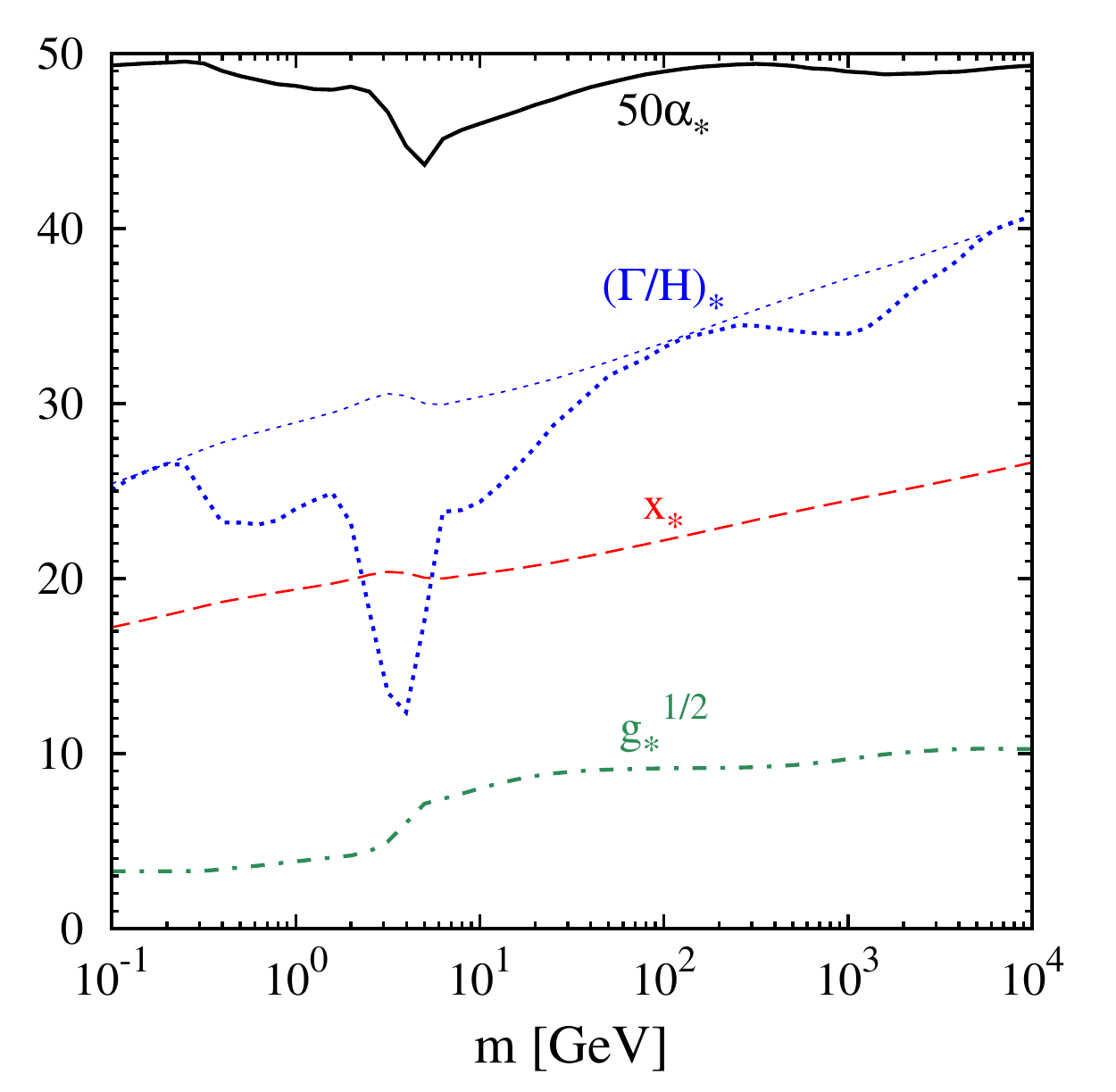}
\caption{The matching point, $x_{*}$ (dashed red), is shown for WIMP masses from 100 MeV to 10 TeV, along with $g_{*}^{1/2}$ (dot-dashed green) and $(\Gamma/H)_{*}$, the ratio of the annihilation rate to the expansion rate evaluated at $T = T_{*}$ without the logarithmic corrections (dotted blue, upper) and with the logarithmic corrections (dotted blue, lower).  Also shown (solid black) is $50\,\alpha_{*}$ (see Eq.\,(\ref{eq:alphadef})).  See the text for details.}
\label{fig:xvsm}
\end{figure}

In Fig.\,\ref{fig:xvsm}, the result for $x_{*}$ is shown by the dashed (red) curve. This has been done iteratively, choosing the value of $\sigv$ required to produce the correct relic abundance $\Omega h^2 = 0.11$. Once $x_{*}$ is found, $T_{*} = m/x_{*}$ is determined and $g_{*} = g(T_{*})$ may be evaluated; $g_{*}^{1/2}$ is shown as a function of the WIMP mass by the dot-dashed (green) curve in Fig.\,\ref{fig:xvsm}.  In Fig.\,\ref{fig:xvsm} we also show the ratio of the annihilation rate to the expansion rate, $(\Gamma/H)_{*}$ as a function of the WIMP mass (the dotted blue curves). We calculated $(\Gamma/H)_{*}$ at two different levels of accuracy. First, $(\Gamma/H)_{*}$ was calculated assuming the logarithmic changes in $g$ to be negligible. This allowed us to rewrite Eq.\,(\ref{eq:deltagen}) as $(\Gamma/H)_{*}=(1+\Delta_*)(x_*-3/2)$. This result is plotted as the upper curve. We calculated a more precise result by including the effect of $d({\rm ln}\,g)/d({\rm ln}\,T)$, which is shown by the lower curve. Note that $(\Gamma/H)_{*}$ is much larger than 1, meaning that when $x = x_{*}$, the annihilation rate far exceeds the expansion rate and significant annihilations occur for $x \gsim x_{*}$; freeze out does not occur when $x = x_{*}$.

\subsubsection{Approach to Freeze-Out}
\label{sec:freeze}

For $x > x_{*}$ (for temperatures $T < T_{*}$), $\Delta$ increases rapidly (exponentially) so that $Y \gg Y_{eq}$, greatly simplifying the evolution equation to
\beq
{dY \over dx}= -{s\sigv \over Hx}\bigg[1 + \frac{1}{3}\frac{d({\rm ln}\,g)}{d({\rm ln}\,T)}\bigg] Y^{2}\,.
\eeq
This equation can be integrated from $x=x_*$ to freeze out $x=x_f$,
\beq
\int_{Y_*}^{Y_f}\,{dY \over Y^2} = - \int_{x_*}^{x_f}\,{dx}{s\sigv \over Hx}\left[1 + \frac{1}{3}\frac{d({\rm ln}\,g)}{d({\rm ln}\,T)}\right].
\eeq
Using $s\sigv/(Hx)\propto\sqrt{g}/x^2$,
\beq
{Y_f \over Y_*}= {1 \over 1 + \alpha_{*}\left(\Gamma/H\right)_{*}}\,,
\label{eq:yfovys}
\eeq
where
\beq
\alpha_*\equiv\int_{T_f}^{T_*}{dT\over T_*}\,\sqrt{g \over g_{*}}\bigg[1 + {1 \over 3}{d({\rm ln}\,g) \over d({\rm ln}\,T)}\bigg]\,.
\label{eq:alphadef}
\eeq
The integral $\alpha_*$ includes the effect of the changing values of $g(T)$ and can be evaluated numerically. Although, strictly speaking, $T_f$ should  be taken to be the present temperature, we evaluate it by assuming that $T_f=T_*/100$ (most of the contribution to the integral comes from $T_{*}/2 \lsim T \lsim T_{*}$). In Fig.\,\ref{fig:xvsm}, $\alpha_*$ is shown by the solid (black) curve, multiplied by 50 for legibility, as a function of WIMP mass.

It should be emphasized that in this analysis the relic abundance does not freeze out when $T = T_{*}$.  Ongoing annihilations between $T=T_{*}$, where $(\Gamma/H)_{*} \gg 1$, and freeze out at temperature $T=T_{f}$, where $(\Gamma/H)_{f} \ll 1$, further reduce the WIMP abundance by the large factor $1 + \alpha_{*}(\Gamma/H)_{*} \gg 1$ (see the dotted blue curves in~Fig.\,\ref{fig:xvsm}), with most of the residual annihilations occurring for~\mbox{$T_{*} \geq T \gsim T_{*}/2$}. Thus, it is expected that the value of $(\Gamma/H)_{*}$ will have an impact on the predicted relic density. Note that previous studies have ignored the $1$ in the denominator of Eq.\,(\ref{eq:yfovys}) and have assumed that $\alpha_*=1$. These approximations incur an error of $\sim 3-5\,\%$ and can affect the calculation substantially, especially for masses in the range $1-10\,$GeV, where the impact of the changing values of $g(T)$ is large. As may be seen from Fig.\,\ref{fig:xvsm}, both $(\Gamma/H)_{*}$ and $\alpha_*$ depend strongly on mass. Our analytical framework takes these effects into account.

\subsubsection{Relic Abundance}
 \label{sec:relic}
 
Having determined $Y_f$, (see Eq.\,(\ref{eq:yfovys})), calculating the relic abundance is straightforward. The frozen out WIMP abundance $Y_f$ is equal to the present day WIMP abundance ($Y_{f} = Y_{0}$), so that the cosmological WIMP mass fraction is
\bea
\Omega&=&{m\,Y_f\,s_0 \over \rho_{crit}}\nonumber\\
&=&{8\pi G \over 3H_0^2}\bigg({mH_*s_0 \over\sigv s_*}\bigg)\bigg({(\Gamma/H)_* \over 1 + \alpha_*(\Gamma/H)_*}\bigg)\,,
\eea
resulting in
\beq
\Omega h^{2} = {9.92\times10^{-28} \over \sigv }\,\bigg({x_{*} \over g_{*}^{1/2}}\bigg)\bigg({(\Gamma/H)_* \over 1 + \alpha_*(\Gamma/H)_*}\bigg).
\label{eq:sigvomegahsq}
\eeq

Note that this result has no explicit mass dependence but $x_{*},\, g_{*},\, {\rm and}\, \alpha_*,$ and $(\Gamma/H)_{*}$ are all mass-dependent.  Recall that the units for units for $\sigv$, here and elsewhere, are ${\rm cm^3s^{-1}}$. For $10^{-1} \leq m{\rm\,(GeV)} \leq 10^4$ we find that $0.97 \lsim (\Gamma/H)_{*}/(1+\alpha_*(\Gamma/H)_{*}) \lsim 1.07$, varying noticeably with mass, as shown in Fig.\,\ref{fig:xvsm}. In most previous analyses the term involving $(\Gamma/H)_{*}$ in Eq.\,(\ref{eq:sigvomegahsq}) is either ignored or assumed to be unity. This small but non-negligible effect is relevant for the low mass regime, that is currently of great interest, and retaining it we find
\beq
10^{26}\sigv= 0.902\bigg({0.11\over\Omega h^2}\bigg)\bigg({x_{*} \over g_{*}^{1/2}}\bigg)\bigg({(\Gamma/H)_* \over 1 + \alpha_*(\Gamma/H)_*}\bigg)\,.
\label{eq:0.11}
\eeq
This result for $\sigv$ as a function of the WIMP mass, assuming the a best-fit value for $\Omega h^2=0.11$, is shown as the dashed (red) curve in Fig.\,\ref{fig:sigvm}. This general result for the relic abundance of a thermal WIMP, whether or not it is a dark matter candidate, derived by an approximate analytic approach to solving the evolution equation\,\cite{zeld65,Steigman:1979kw} agrees to better than $\sim 3\,\%$ with the results of the direct numerical integration of the evolution equation (solid black curve in Fig.\,\ref{fig:sigvm}) described below in \S\ref{sec:numerical}. For analytic results accurate to $\sim 5\,\%$, the last factor in Eq.\,(\ref{eq:0.11}) may be approximated by 1.02.

\subsection{Numerical Results and Discussion}
\label{sec:numerical}

\begin{figure}[!t]
\includegraphics*[width=1.0\columnwidth]{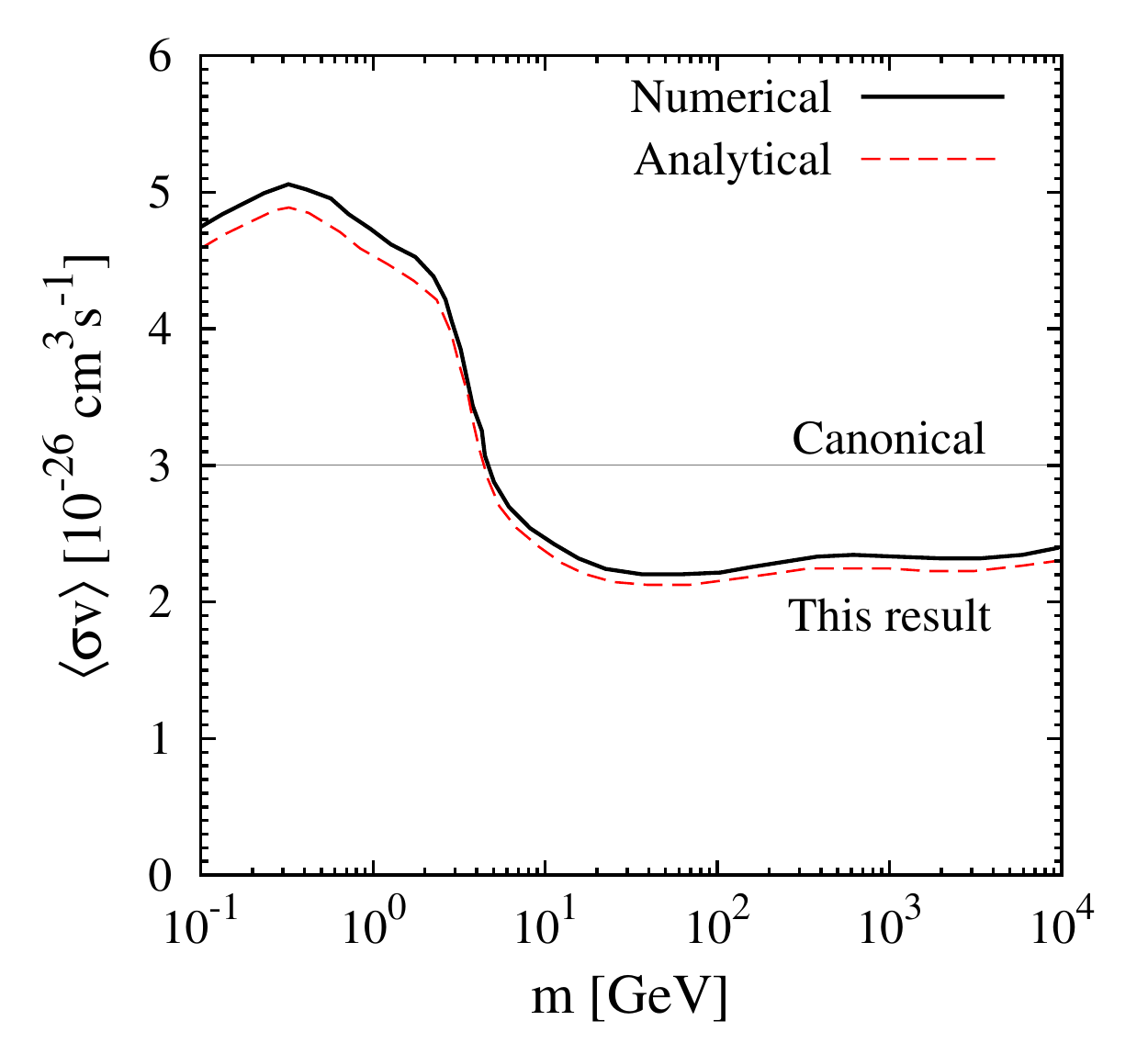}
\caption{The thermal annihilation cross section required for $\Omega_{\chi}h^{2} = 0.11$ as a function of the mass for a Majorana WIMP. The solid (black) curve is from numerical integration of the evolution equation and the dashed (red) curve is for the approximate analytic solution in Eq.\,(\ref{eq:sigvomegahsq}).  Note that the agreement between analytical and numerical results is better than $\sim3\%$. For comparison, the thin horizontal line shows the canonical value $\sigv=3\times10^{-26}\,{\rm cm^3s^{-1}}$.}
\label{fig:sigvm}
\end{figure}

To compare with the approximate analytic results we have calculated the relic abundance by numerically integrating the WIMP evolution equation, Eq.\,(\ref{eq:workingeq}). We transform this equation into a simple dimensionless form,
\beq
{dY \over dx} = \lambda\bigg[1 + {1 \over 3}{d({\rm ln}g_s) \over \,d({\rm ln}T)}\bigg]{g_s \over g_{\rho}^{1/2}}{1\over x^{2}}(Y_{eq}^{2} - Y^{2}),
\eeq
where $\lambda \equiv 2.76\times10^{35}m\sigv$ and $Y_{eq}=0.145\,(g_{\chi}/g_s)\,x^{3/2}{\rm e}^{-x}$ ($m$ is in GeV and $\sigv$ in ${\rm cm^3s^{-1}}$). An approximation made here is to use the non-relativistic expression for $n_{eq}$ in $Y_{eq}$. This has negligible impact on our results. For $m$ in the range $10^{-1} - 10^{\,4}$ GeV and $\sigv$ in the range $10^{-26}-10^{-25}\,{\rm cm^{3}s^{-1}}$, $\lambda$ has values in the range $10^8-10^{14}$. The equation to be integrated is therefore numerically stiff.  We find it useful to make the replacement $W={\rm ln}Y$ and to integrate
\beq
{dW \over dx}={\lambda \over x^2}\bigg[1 + {1 \over 3}{d({\rm ln}g_s) \over \,d({\rm ln}T)}\bigg]{g_s \over g_{\rho}^{1/2}}({\rm e}^{(2W_{eq}-W)}-{\rm e}^{W})\,,
\label{eq:Wevoleq}
\eeq
where $W$ does not change by many orders of magnitude over the range of integration. This significantly reduces the computational effort. In particular, one can work with lower precision and still determine the solution quite accurately.

We integrated Eq.\,(\ref{eq:Wevoleq}) from $x = 1$ to $x = 1000$ using {\tt Mathematica} along with our own numerical routine employing a stiffness switching explicit-implicit method, for $10^{6}$ equally distributed values of $\lambda$ in the $\{\sigv, {\rm log}\,m\}$ plane. After finding $W$ at $x = 1000$, we transform back to $Y$, allowing us to find $\Omega h^{2}$.  After finding the values of $\Omega h^{2}$ corresponding to these $10^{6}$ points, we identify the contour in the $\{\sigv, m\}$ plane corresponding to the choice $\Omega h^{2} = 0.11$.

In Fig.\,\ref{fig:sigvm}, the numerically integrated value of $\sigv$ is shown as a function of mass (solid black curve), verifying our claim that the approximate analytic results agree with the numerical results to better than $3\,\%$ over this range of WIMP masses. Most of the few percent systematic downward shift of the analytic results can be traced to the underestimate of $x_*$ compared to the numerial result shown in Fig.\,\ref{fig:approx}. Bender and Sarkar have recently solved the relic evolution equation using boundary-layer theory~\cite{Bender:2012gc}, and the asymptotic solution in their Eq.\,(46) predicts that the required $\sigv$ is larger by a factor of $(x_*+1)/x_*$ (compared to our solution). This correction brings the analytical results to even closer agreement ($\sim1\%$) with the numerical results. Additionally, we find that varying $g(T)$ within its uncertainties changes the numerical results by only $\sim1\,\%$. This underscores our expectation that the relic abundance can now be calculated quite precisely, with all uncertainties constrained to be quite small. We now proceed to describe our results, and ascertain their impact on WIMP annihilation searches.

We call attention to the result that for low masses the cross section required to account for the observed relic dark matter density is mass dependent, reaching a maximum of $\sigv \approx 5.2\times 10^{-26}\,{\rm cm^{3}s^{-1}}$ for $m \approx 0.3$~GeV.  As the WIMP mass increases from this value, $\sigv$ first decreases by more than a factor of two, reaching a minimum at $\approx 2.2\times 10^{-26}\,{\rm cm^{3}s^{-1}}$ when $m \approx 30$~GeV, and then $\sigv$ begins a slow increase to $\approx 2.4\times 10^{-26}\,{\rm cm^{3}s^{-1}}$ for $m \approx 10$~TeV.  The exact shape of the rise of the cross section at low mass depends on the quark masses and the temperature of the quark-hadron transition~\cite{Laine:2006cp}. To the best of our knowledge, this rise in $\sigv$ for low WIMP masses has only been noted in some specific supersymmetric WIMP models, \eg,~\cite{Bottino:2003iu, Cerdeno:2011tf}. Kappl and Winkler~\cite{Kappl:2011jw} also plot a similar feature, with much larger uncertainties, in their Fig.\,4. Numerical packages, \eg, {\tt DarkSUSY}~\cite{Gondolo:2004sc}and {\tt micrOMEGAs}~\cite{Belanger:2006is}, do reproduce this effect, but previous analytical calculations have ignored it. We emphasize that this is a generic feature and make it manifestly visible in our analytical results. The rise at low mass is a reflection of the fact that for this mass range the number of relativistic degrees of freedom populated at $T \leq T_{*}$ is changing rapidly (decreasing with decreasing mass) due to the quark-hadron transition, while $x_{*}$ remains roughly constant (see Fig.\,\ref{fig:xvsm}) so that the combination $x_{*}/g_{*}^{1/2}$ increases with decreasing mass.

Over the remaining range from $\sim 10$~GeV to $\sim 10$~TeV, $\sigv = 2.2\times 10^{-26}\,{\rm cm^{3}s^{-1}}$ within $\sim 5\,\%$, which is $\sim 40\,\%$ lower than the canonical value $3\times10^{-26}\,{\rm cm^{3}s^{-1}}$ usually quoted in the literature. The origin for this discrepancy is not completely clear. The often quoted reference by Jungman, Kamionkowski, and Griest~\cite{Jungman:1995df} provides $\sigv\Omega h^{2}\approx3\times10^{-27}{\rm cm^3s^{-1}}$, which may have resulted from the approximate treatment and rounding-off to one significant figure. In fact, Steigman's original calculation~\cite{Steigman:1979kw}, when modified for Majorana WIMPs, gives $\sigv\Omega h^{2}=2.5\times10^{-27}{\rm cm^3s^{-1}}$, which could have been rounded upwards. Our more careful approach yields an answer that is $5\%$ smaller, and agrees better with the value that we find numerically. This additional precision has become relevant only recently due to the accurate determination of $\Omega h^{2}$ and experimental sensitivity having approached this thermal scale.

\subsection{Summary of Results for $\sigv$ versus $\Omega h^{2}$}

From Eq.\,(\ref{eq:sigvomegahsq}) and the discussion below it in \S\ref{sec:relic}, the connection between $\sigv$ and $\Omega h^{2}$, accurate to $\sim 5\%$ or better, is
\beq
10^{27}\sigv\Omega h^{2} = 1.0(x_{*}/g_{*}^{1/2}).
\eeq
As may be seen from Figs.\,\ref{fig:xvsm} and \ref{fig:sigvm}, for $m \gsim 10$\,GeV, the ratio of $x_{*}$ to $g_{*}$ is very nearly independent of mass, resulting in $10^{27}\sigv\Omega h^{2} \approx 2.4$.  However, as may be seen from Fig.\,\ref{fig:evol}, and from Eq.\,(\ref{eq:x*vsm}), even for this mass range $x_{*}$ and $g_{*}$ do depend, logarithmically, on $\sigv$ and $m$.  For $m \gsim 10$\,GeV we have found that as $\sigv$ varies over 12 orders of magnitude, from $\sigv \sim 10^{-27}\,{\rm cm^{3}s^{-1}}$ to $\sigv \sim 10^{-15}\,{\rm cm^{3}s^{-1}}$, the connection between $\sigv$ and $\Omega h^{2}$, as a function of $\sigv$, is well fit by
\beq
10^{27}\sigv\Omega h^{2} = 2.0 + 0.3\,{\rm log}(10^{27}\sigv).
\eeq
If, instead, $\Omega h^{2}$ is known, then the same relation, as a function of $\Omega h^{2}$, is well fit by
\beq 
10^{27}\sigv\Omega h^{2} = 2.1 - 0.3\,{\rm log}(\Omega h^{2}).
\eeq
For $m \gsim 10$\,GeV and the current best estimate of $\Omega h^{2} = 0.11$, the required annihilation cross section is $\sigv = 2.2\times 10^{-26}\,{\rm cm^{3}s^{-1}}$, within $\sim 5\%$.  Because of the rapid change in $g_{*}$ for $T_{*}$ in the vicinity of the quark hadron transition temperature, there is no correspondingly simple $\sigv$ versus $\Omega h^{2}$ relation for lower WIMP masses.


\subsection{Variations on a Theme}

So far we have confined our discussion to the simplest WIMP scenario in a standard cosmological setting. We now discuss some possible generalizations of and exceptions to our results. First, we have assumed the WIMP to be a Majorana fermion $\chi$, so that $\chi$ and $\bar{\chi}$ are indistinguishable.  In contrast, if $\chi$ is a Dirac fermion $\chi \neq \bar\chi$, and $g_{\chi}$ is doubled. In this case, for particle-antiparticle symmetry, the relic abundance of $\chi$ is half as large. This merely increases the required value of $\sigv$ by a factor of 2 for Dirac fermions.

In our calculations the particle content of the standard model has been assumed.  This only enters through $g(T)$. In addition, we have assumed that the WIMP is the only new (non-SM) particle. We note that CMB analyses which allow for ``extra" relativistic degrees of freedom (``equivalent neutrinos") favor slightly higher dark matter densities, $\Omega_{\rm CDM}h^{2} \approx 0.13 - 0.14$\,\cite{Komatsu:2010fb}.  These lead to correspondingly lower values of $\sigv$, by $\sim 20-30\,\%$.  The presence of such extra, decoupled, relativistic degrees of freedom ($\Delta{N}_{\nu}$) will modify the analysis presented here in a manner that is model dependent (How many extra degrees of freedom? When did they decouple?)\,but, for $\Delta{N}_{\nu} \lsim 2$, the higher CDM mass density inferred from the CMB dominates and, since $\sigv \propto (\Omega h^{2})^{-1}$, a smaller annihilation cross section is required. For non-standard models containing additional decoupled extremely relativistic particles, the distinction between $g_{\rho}$ and $g_{s}$ may need to be taken into account. This is model dependent, but may be included relatively easily. 

The s-wave dominated annihilation cross section $\sigv$ is independent of temperature.  It is straightforward to generalize our results to p-wave (or arbitrary \mbox{{\it l\,}-wave}) annihilation. Simply rewriting $\sigv\rightarrow\sigv_0x^{2l}$, our analysis can be repeated for analogous results, introducing no additional errors. However, for anything but s-wave annihilations, the present experimental constraints on annihilation are not even close to probing the relevant thermal scale.

A major assumption in our analysis is that entropy is conserved throughout the relevant evolution of the Universe. This assumption is justified at the quark-hadron transition, which is a crossover transition\,\cite{Brown:1988qe, Aoki:2006we, Gupta:2011wh} generating no entropy. However, we have assumed that the electroweak transition is at most a second order phase transition generating no entropy.  If the electroweak transition is first order, accompanied by an inflationary period, the calculation of the relic WIMP abundance will depend on the ``reheat" temperature ($T_{RH}$), and thermal relics will be absent or suppressed if $T_{RH} < m_{\chi}$.

WIMP annihilation may be more complicated than the simple picture adopted in our analysis. There may be other particles almost degenerate in mass with the WIMP that contribute to the relic annihilation  (coannihilation) rate, or there may be effects due to mass thresholds, or resonances~\cite{Griest:1990kh}. These effects are model dependent, and we have no way of easily generalizing our results. More detailed numerical analyses for specific models is needed in such scenarios. A non-standard cosmological expansion or a non-thermal dark matter candidate may require a separate treatment~\cite{Feng:2010tg}.

\section{Confronting Experimental Limits}
\label{sec:limits}

The program for the indirect detection of DM is to search systematically for annihilation fluxes into all channels at all possible energies. Given a WIMP mass, only a limited number of final states lighter than the WIMP are kinematically allowed because the annihilating WIMPs are non-relativistic today. Existing experiments have recently reached the sensitivity to probe the thermal relic annihilation cross sections for some channels. We now discuss the impact of our results on these recent indirect detection constraints on WIMP models using gamma rays observed by Fermi-LAT and from cosmology using the CMB observations of WMAP+ACT.  We also briefly comment on the impact our results have on the interpretation of direct detection experiments.

\begin{figure}[!t]
\includegraphics[width=1.0\columnwidth]{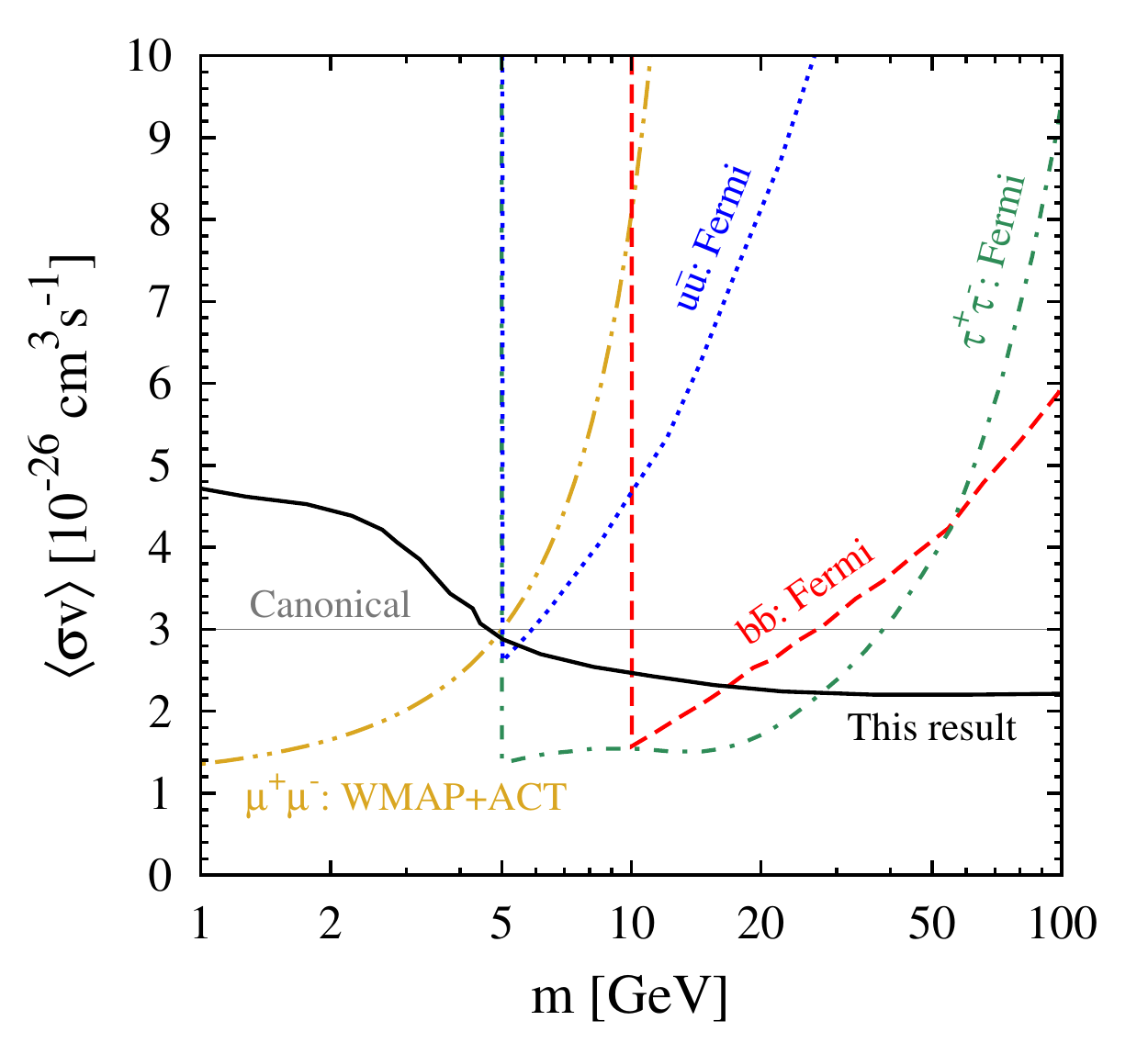}
\caption{Comparison of the Fermi-LAT limits on the thermal WIMP annihilation cross section into particular channels (the regions above the colored curves are ruled out at $95\,\%$ c.l.) with the relic annihilation cross section calculated for $\Omega h^{2} = 0.11$ (solid black). Constraints from the Milky Way for $u\bar{u}$ (dotted blue); constraints from dwarf galaxies for $b\bar{b}$ (dashed red), and $\tau^+\tau^-$ (dot-dashed green); constraints from cosmology for $\mu^+\mu^-$ (dot-dot-dashed yellow).}
\label{fig:fermi}
\end{figure}

In Fig.\,\ref{fig:fermi}, we plot the digitally extracted data on the limits to the WIMP annihilation cross section from (\emph{i}) Analysis of the diffuse gamma ray flux in Milky Way~\cite{Abazajian:2010sq, Abdo:2010dk}, (\emph{ii}) Stacked analysis of the gamma ray flux from $10$ dwarf spheroidal satellite galaxies of the Milky Way by the Fermi-LAT~\cite{GeringerSameth:2011iw, collaboration:2011wa}, and (\emph{iii}) Constraints from reionization and recombination based on an analysis of WMAP+ACT data~\cite{Cirelli:2009bb,Hutsi:2011vx,Galli:2011rz}. The strongest limits from the diffuse flux analysis are for the $u\bar{u}$ annihilation channel~(dotted blue curve). The $d\bar{d}$ limits, which are similar, are not shown here. Other channels give somewhat weaker limits. The data from dwarf galaxies provides a stronger set of bounds for the $b\bar{b}$ (dashed red curve) and $\tau^+\tau^-$ (dot-dashed green curve) channels. For the lowest masses, the constraints from cosmology (dot-dot-dashed yellow curve) are the strongest, but they are less direct than the gamma ray observations.

Taking the diffuse flux results at face value and if the canonical value of the thermal cross section, \ie, $\sigv = 3\times 10^{-26}\,{\rm cm^{3}s^{-1}}$, were used, WIMP masses in the $5-7\,$GeV range suggested by the direct detection experiments would be disfavored for annihilations resulting mainly in the light $u\bar{u}$ quarks (compare the dotted blue and horizontal grey curves). The limits from annihilation to heavier quarks and leptons are weaker. However, these constraints weaken when the lower value of $\sigv$ based on our analysis (solid black curve) is used.  More interesting, the rise in the annihilation cross section for lower masses suggests that if the Fermi-LAT analysis were extended to slightly lower gamma ray energies, corresponding to WIMP masses in the $1-5$\,GeV range, they could provide stronger constraints compared to those inferred using the canonical cross section.

The stacked dwarf galaxy analysis provides mass limits for annihilations to the $b\bar{b}$ and $\tau^+\tau^-$ channels, improving the diffuse flux limits constraints by almost an order of magnitude. It is interesting to note that for the WIMP mass range $\gsim5\,$GeV, the $\tau^+\tau^-$ constraint already rules out $\tau^+\tau^-$ branching fractions larger than $\sim50\,\%$. In particular, WIMP masses in the range $5-27\,$GeV are ruled out for the $\tau^+\tau^-$ channel, and in the range $10-17\,$GeV, for the $b\bar{b}$ channel (compare the dot-dashed green or dashed red curves, respectively, to the solid black curves). These mass limits are a factor of $\sim2$ weaker than those by Geringer-Sameth and Koushiappas~\cite{GeringerSameth:2011iw} or the Fermi-LAT collaboration~\cite{collaboration:2011wa}, as a direct result of the nearly $40\,\%$ reduction in the thermal annihilation cross section pointed out in this paper.

Note that for the above limits from the diffuse flux and from dwarf galaxies, the analyses have been limited to higher energy gamma rays, resulting in the sharp cut-offs to the limits at low masses shown in Fig.\,\ref{fig:fermi}. In general, these cut-offs are above the kinematic thresholds for the corresponding channels. Usually, lowering the threshold would have no advantage and would simply lead to worse detector performance.  In this case however, the larger cross sections at lower masses that we have pointed out here should make it easier to extend the gamma ray analyses to lower energies, corresponding to smaller WIMP masses, where, although the backgrounds are higher, so too is the expected flux from WIMP annihilation.

Cosmological constraints from reionization and recombination disfavor low mass WIMPs~\mbox{\cite{Cirelli:2009bb,Hutsi:2011vx,Galli:2011rz}}. If the DM is a thermal WIMP, as we have assumed, our results here imply that the cosmological constraints are stronger than those using the canonical value of $\sigv$ (compare the dot-dot-dashed yellow curve to the solid black curve, instead of the horizontal grey curve). These constraints may, however, be evaded if the annihilation is primarily to neutrinos, where the limits are weaker~\cite{Beacom:2006tt, Yuksel:2007ac, Yuan:2010gn, Abbasi:2011eq}. In addition, the cosmological constraints are indirect and depend on different assumptions.

Direct detection experiments, \eg, DAMA/LIBRA, CoGeNT, CRESST-II etc.~\cite{Bernabei:2010mq, Aalseth:2011wp,Aalseth:2010vx,Angloher:2011uu}, which involve WIMP scattering on nuclei prefer the $1-10\,$GeV region, but relating the WIMP scattering cross section to the annihilation cross section is model dependent. However, for any given model the annihilation cross section can be related to the scattering cross section, \eg,~\cite{Cao:2009uv,Keung:2010tu}. It is important to note that models designed to explain the CoGeNT and DAMA results must now do so without exceeding the branching fraction for annihilation into $u,\,d,$ quarks allowed by the above constraints which follow from the annihilation cross section required to account for the thermal relic abundance. These results can also be interpreted as setting an upper limit to the allowed branching ratio for annihilations to the light quark channels.  This will set constraints on the detailed particle physics models.  For example, we find that for the models considered by Keung\,\etal~\cite{Keung:2010tu}, CoGeNT prefers values of $\sigv$ which are in the $10^{-27} - 10^{-25}\,{\rm cm^3s^{-1}}$ range. Therefore, a large fraction of the non-universal scalar models considered in~\cite{Keung:2010tu} are disfavored as thermal~DM.

Gamma ray fluxes from dark matter annihilation in clusters of galaxies are expected to be too small to be detected by $\sim 2-3$ orders of magnitude, but substructure of very high density on scales smaller than those observed, or normally probed by N-body simulations, could significantly enhance the annihilation rate, increasing the resulting gamma ray flux dramatically.  Adopting such model dependent, small scale halo structure in the analyses of the flux of gamma rays from nearby galaxy clusters leads to stringent limits to $\sigv$~\cite{Ackermann:2010rg, Ando:2012vu}. If the possible substructure is included according to Han \etal~\cite{Han:2012au}, the limits in the $\mu^+\mu^-$ and $b\bar{b}$ channels may now be below the total annihilation cross section needed to reproduce the relic abundance of WIMPs for WIMP masses in the range $\sim 5-40$~GeV. 

The recent frenzy of activity suggesting values of $\sigv$ close to that predicted for a thermal relic, along with the prospect of new gamma ray data, provided the stimulus for our revisiting the relic abundance analysis and quantifying the approximations and uncertainties. It is clear that the available limits need to be interpreted carefully. Astrophysical uncertainties on these constraints can be as large as an order of magnitude, but the precise quantitative relation between $\sigv$ and the relic abundance can also have a strong impact on conclusions inferred from them. In the future, these analyses will be extended to all possible channels (particularly to those involving lighter quarks and leptons), and the limits will be combined to obtain a lower bound to the WIMP mass, assuming that it annihilates into observable channels, and no signal is observed.  In particular, the rise in $\sigv$ at low masses noted here suggests that annihilation into, \eg, $u\bar{u}$ and $d\bar{d}$, can be probed more easily than would be expected for the canonical value of $\sigv$. This should provide motivation to revisit many analyses and to extend them to lower energies, in order to probe low mass WIMPs.
 
\section{Outlook and Conclusions}
\label{sec:summary}

A key result of the work described here is to point out that the thermal cross section, $\sigv$, required to account for the relic dark matter abundance can be calculated with great precision $(\lsim {\rm few}\,\%)$. In addition, we find that $\sigv$ is not independent of the WIMP mass for masses $\lsim 10$~GeV, as a result of the relatively rapid decrease in the number of relativistic degrees of freedom for temperatures below the quark-hadron transition.  From our more careful calculation we find that while the required cross section is very nearly independent of mass for larger WIMP masses, and that it is $\sim 40\,\%$ smaller than the canonical value of $\sigv$ usually adopted in the literature.  While these differences may seem modest, as shown above in Fig.\,\ref{fig:fermi}, they do have a noticeable quantitative impact on the interpretation of recent results from various experiments.  When relic abundances are calculated in more detailed particle physics models of dark matter the mass dependence is taken into consideration so, as a matter of principle, the mass dependence for the generic WIMP model should be included as well. Our analysis does that.

We expect that analyses of future gamma ray data from the Milky Way, dwarf galaxies, and clusters are likely to yield even more stringent constraints on DM models. Future results from Fermi-LAT and from cosmology will have the potential to probe the low mass region even more  aggressively by analyzing annihilation into various light particle final states, so that the small differences, that we have pointed out, are likely to become even more relevant.

\section*{Acknowledgments}
We thank M.\,Laine and Y.\,Schroeder for helpful clarifications and for sharing with us their calculated values of $g(T)$.  We also acknowledge useful discussions with R.\,Laha and R.\,J.\,Scherrer. The research of G.\,S.~and B.\,D. is supported by DOE Grant DE-FG02-91ER40690.  The research of B.\,D.\, is also supported by a CCAPP Fellowship.  The research of J.\,F.\,B. is supported by NSF Grant PHY-1101216.  


\end{document}